\input quarkonium.sty
%A4 size
\def\dinafour{
\setdimensions{17cm}{24.2cm}}
\dinafour
\magnification1000
\brochureb{\smallsc a. pineda and f. j.  yndur\'ain}{\smallsc calculation of quarkonium 
spectrum and $m_b,\,m_c$ to order $\alpha_s^4$}{1}
\rightline{UB-ECM-PF 97/34}
\rightline{FTUAM 97-15\qquad}
\rightline{November, 11, 1997}
\bigskip
\hrule height .3mm
\vskip.6cm
\centerline{{\bigfib Calculation of Quarkonium Spectrum and $m_b,\,m_c$
 to Order $\alpha_s^4$}\footnote*{\petit Supported in part by CICYT, Spain}}
\medskip
\centerrule{.7cm}
\vskip1cm
\setbox8=\vbox{\hsize65mm {\noindent\fib A. Pineda} 
\vskip .1cm
\noindent{\addressfont Departament d'Estructura i Constituents 
de la Mat\`eria,\hb
 Universitat de Barcelona,\hb
 Avinguda Diagonal, 647\hb
E-08208, Barcelona, Spain.}}
\centerline{\box8}
\smallskip
\setbox7=\vbox{\hsize65mm \fib and} 
\centerline{\box7}
\smallskip
\setbox9=\vbox{\hsize65mm {\noindent\fib F. J. 
Yndur\'ain} 
\vskip .1cm
\noindent{\addressfont Departamento de F\'{\i}sica Te\'orica, C-XI,\hb
 Universidad Aut\'onoma de Madrid,\hb
 Canto Blanco,\hb
E-28049, Madrid, Spain.}\hb}
\smallskip
\centerline{\box9}
\bigskip
\setbox0=\vbox{\abstracttype{Abstract}We include two loop, relativistic one loop and second order
 relativistic tree level corrections, plus leading 
 nonperturbative contributions, to obtain a calculation of the lower states in the heavy
 quarkonium spectrum correct 
up to, and including, $O(\alpha_s^4)$ and leading $\Lambdav^4/m^4$ terms. This 
allows us, in particular, to obtain a model independent determination
 of the pole masses of the $b,\,c$ quarks, 
$$m_b=5\,015^{+110}_{-70}\;\mev;\;
m_c=1\,884^{+222}_{-133}\;\mev$$
to which correspond the $\overline{\hbox{MS}}$ masses,
$$\bar{m}_b(\bar{m}_b^2)=4\,453^{+50}_{-32}\;\mev;\;
\bar{m}_c(\bar{m}_c^2)=1\,547^{+169}_{-102}\;\mev.$$

The decay $\Gammav(\Upsilonv\rightarrow e^+e^-)$ is found in agreement 
with experiment,
$$\Gammav(\Upsilonv\rightarrow e^+e^-)=1.135^{+0.27}_{-0.29}\;\kev
\;(\hbox{exp.}=1.320\pm0.04\,\kev),$$
and the hyperfine splitting is predicted to be
$$M(\Upsilonv)-M(\eta)=48.5^{+15.7}_{-12.2}\;\mev.$$}
\centerline{\box0}
\brochureendcover{Typeset with \physmatex}

\brochuresection{1 Introduction}
In recent years it has become possible to perform rigorous QCD analyses of  
heavy quarkonium systems, and this due to two reasons. First,
 radiative corrections have been calculated to increasing order
 of accuracy. The one loop corrections to the nonrelativistic (NR) 
spin-independent (SI) potential were calculated already in 1980\ref{1}.
 This was extended in refs.~2, 3 to the
 spin-dependent corrections, and in ref.~4 by including the velocity
 corrections to the SI part. Finally, the two loop nonrelativistic, 
spin independent  corrections to the potential have
 been evaluated recently\ref{5}.

Secondly, Leutwyler and Voloshin (ref.~6; see also ref.~7)  have shown how to
 take into account, to leading order, nonperturbative (NP) effects, associated
 with the nonzero value of various condensates, of which the leading contribution 
is that of the gluon condensate 
$\langle \alpha_sG^2\rangle$. This has been implemented, together with the 
potential obtained with radiative corrections to one loop, in refs.~4, 8 
where a study of bound states $\bar{b}b$ with $nl=10,\,20,\,21$ and $\bar{c}c$ 
sates with $nl=10$ was given\fnote{Throughout this paper $n,\,l$ will 
denote the principal quantum number and the angular momentum of bound states.}.
 The analysis was extended in refs.~9, 7 with 
the inclusion of size effects and higher condensates.

The overall conclusion of these analyses is that pure QCD, without recourse 
to introducing phenomenological interactions, produces a good description with
 manageable errors of the $\bar{b}b$ ground state and, to a lesser 
extent, of the splitting $M(\Upsilonv)-M(\eta_b)$ and the decay 
$\Upsilonv\rightarrow e^+e^-$. The description of the ground state of 
$\bar{c}c$ and of the excited states $n=2,\,l=0,1$ of $\bar{b}b$ was shown 
to be even less reliable: the corrections are large, in some cases larger
 than the nominally leading terms. Still, it was possible, by using the 
renormalization point $\mu$ as a free parameter, to get a fairly accurate 
description of all $n=2$ states including tensor and LS  splittings\ref{8}.

In the present paper we extend this analysis by including the two loop
 corrections to the SI, NR potential recently calculated\ref{5} adding also 
velocity corrections to certain one loop pieces to get a 
calculation accurate up to, and including, $O(\alpha_s^4)$ corrections. By 
taking into account the leading nonperturbative terms,
 we also include in the analysis corrections of order $\Lambdav^4/m^4$.  
The main improvement so obtained is that we get an extremely stable 
and precise determination of the ground state of $\bar{b}b$ and, to 
a lesser extent, $\bar{c}c$. If we invert the calculations we can 
deduce quark masses from the masses of the $\Upsilonv,\,J/\psi$ particles. 
We then find, for the pole masses,
$$\eqalign{m_b=5\,015^{+110}_{-70}\;\mev,\cr
m_c=1\,884^{+222}_{-133}\;\mev\cr}\equn{(1.1a)}$$
to which correspond the $\overline{\hbox{MS}}$ masses,
$$\eqalign{\bar{m}_b(\bar{m}_b^2)=4\,453^{+50}_{-32}\;\mev,\cr
\bar{m}_c(\bar{m}_c^2)=1\,547^{+169}_{-102}\;\mev.\cr}\equn{(1.1b)}$$ 
 The error includes the estimated theoretical error of the calculation, 
see the text. Note that (1.1) are very precise: as stated they are correct to order 
$\alpha_s^4$, and leading, $O(\Lambdav^4/m^4)$ nonperturbative effects. This 
is to be compared with estimates based on sum rules\ref{10} which are only 
accurate to order $\alpha_s^2$, or previous bound state calculations\ref{4}, accurate 
only to third order in $\alpha_s$.

The decay $\Gammav(\Upsilonv\rightarrow e^+e^-)$ is also 
given in agreement 
with experiment, within errors:
$$\Gammav(\Upsilonv\rightarrow e^+e^-)=1.135^{+0.27}_{-0.29}\;\kev
\;(\hbox{exp.}=1.320\pm0.04\,\kev).$$
The hyperfine splitting is predicted to be
$$M(\Upsilonv)-M(\eta)=48.5^{+15.7}_{-12.2}\;\mev.$$

For higher states ($nl=20,\,21$) the errors are much larger but, 
within these, one has compatibility with experiment (cf. \subsect~4.3).

\brochuresection{2 The Effective Potential}
We follow the method of effective potentials of Gupta et al\ref{3}, and the
 renormalization scheme of ref.~4. The Hamiltonian for quarkonium may 
then be split in the following form:
$$H=H^{(0)}+H_1\equn{(2.1a)}$$
where $H^{(0)}$ is
$$\eqalign{H^{(0)}=&2m+\dfrac{-1}{m}\lap-\dfrac{C_F\tilde{\alpha}_s(\mu^2)}{r},\cr
\tilde{\alpha}_s(\mu^2)=&\alpha_s(\mu^2)
\left\{1+\left(a_1+\dfrac{\gammae\beta_0}{2}\right)\dfrac{\alpha_s(\mu^2)}{\pi}\right.\cr
&\left.+\left[\gammae\left(a_1\beta_0+\dfrac{\beta_1}{8}\right)+
\left(\dfrac{\pi^2}{12}+\gammae^2\right)\dfrac{\beta_0^2}{4}+
b_1\right]\dfrac{\alpha_s^2}{\pi^2}\right\}\cr}\equn{(2.1b)}$$
and can (and will) be solved exactly. $H_1$ may be written as
$$H_1=V_{\rm tree}+V^{(L)}_1+V^{(L)}_2+V^{(LL)}+V_{\rm s.rel}+V_{\rm spin},\equn{(2.1c)}$$
and
$$\eqalign{V_{\rm tree}=&\dfrac{-1}{4m^3}\lap^2+\dfrac{C_F\alpha_s}{m^2r}\lap,\cr
V^{(L)}_1=&\dfrac{-C_F\beta_0\alpha_s(\mu^2)^2}{2\pi}\,\dfrac{\log r\mu}{r},\cr
V^{(L)}_2=&\dfrac{-C_F\alpha_s^3}{\pi^2}\,
\left(a_1\beta_0+\dfrac{\beta_1}{8}+\dfrac{\gammae\beta_0^2}{2}\right)\dfrac{\log r\mu}{r}\cr
&\equiv\dfrac{-C_Fc_2^{(L)}\alpha_s^3}{\pi^2}\;\dfrac{\log r\mu}{r},\cr
V^{(LL)}=&\dfrac{-C_F\beta_0^2\alpha_s^3}{4\pi^2}\;\dfrac{\log^2 r\mu}{r},\cr
V_{\rm s.rel}=&\dfrac{C_Fa_2\alpha_s^2}{2mr^2},\cr
V_{\rm spin}=&\dfrac{4\pi C_F\alpha_s}{3m^2}s(s+1)\delta({\bf r}).\cr}\equn{(2.1d)}$$
Here the running coupling constant has to be taken to three loops. For
 the values of the constants entering above formulas, cf. the Appendix.
 $a_1$ was calculated in 
ref.~1, $a_2$ in ref.~4 and $b_1$ in ref.~5. The other terms in (2.1d) can 
be obtained by use of the renormalization group, see e.g. ref.~4, or are well-known 
tree level relativistic corrections (including kinetic energy corrections).

A few words are due on \equs~(2.1). First of all, they only take into account the 
{\sl perturbative} part of the calculation; NP effects will be incorporated later.
 Secondly, it should be noted that $H_1$ contains a velocity-dependent one
loop piece, $V_{\rm s.rel}$. This is because the average velocity in a Coulombic potential 
is $\langle |v|\rangle\sim \alpha_s$, hence 
a calculation correct to order $\alpha_s^4$ requires tree level $O(v^2)$ and 
one loop $O(|v|)$ contributions. 
All these terms in $H_1$ may be treated as perturbations to first order, {\sl except} $V^{(L)}$. 
For this, the second order perturbative contribution is required as this
 also produces a correction of order $\alpha_s^4$.

A last comment concerns the renormalization scheme. We have followed ref.~4 
in renormalizing $\alpha_s$ in the \msbar\ scheme; but the mass $m$ that
 appears in \equs~(2.1) is the two loop pole mass. That is to say, it is defined by the 
equation,
$$S^{-1}_2(\slash{p}=m,m)=0\equn{(2.4)}$$
where $S_2(\slash{p},m)$ is the quark propagator to two loops. One can relate $m$ 
to the \msbar\ parameter, also to two loop accuracy, using the results of refs.~11:
$$\eqalign{\bar{m}(\bar{m}^2)=m\left\{1+\dfrac{C_F\alpha_s(m^2)}{\pi}+
(K-2C_F)\left(\dfrac{\alpha_s}{\pi}\right)^2\right\}^{-1},\cr
K(n_f=4)\simeq13.4;\;K(n_f=3)\simeq14.0 .\cr}\equn{(2.5)}$$
\goodbreak
\brochuresection{3 Energy Shifts}
\brochuresubsection{3.1 Order $\alpha_s^4,\,\Lambdav^4/m^4$}
Taking into account the expression for the Hamiltonian, \equn{(2.1)}, we can write
$$E_{nl}=2m-m\dfrac{C_F^2\tilde{\alpha}_s^2}{4n^2}+\sum_{V}\delta^{(1)}_{V}E_{nl}
+\delta^{(2)}_{V_1^{(L)}}E_{nl}+\delta_{\rm NP}E_{nl}.\equn{(3.1)}$$
Here the $\delta^{(1)}_{V}E_{nl}$ may be easily evaluated with 
the formulas in the Appendix to ref.~4.
 We define generally the analogue of the Bohr radius,
$$a(\mu^2)=\dfrac{2}{mC_F\tilde{\alpha}_s(\mu^2)},$$
and then,
$$\delta^{(1)}_{V_{\rm tree}}E_{nl}=-\dfrac{2}{n^4\,m^3\,a^4}
\left[\dfrac{1}{2l+1}-\dfrac{3}{8n}\right]+
\dfrac{C_F\alpha_s}{m^2}\,\dfrac{2l+1-4n}{n^4(2l+1)a^3};\equn{(3.2a)}$$
$$\delta^{(1)}_{V^{(L)}_1}E_{nl}=
-\dfrac{\beta_0C_F\alpha^2_s(\mu^2)}{2\pi n^2a}
\left[\log\dfrac{na\mu}{2}+\psi(n+l+1)\right];\equn{(3.2b)}$$
$$\delta^{(1)}_{V_2^{(L)}}E_{nl}=-\dfrac{C_Fc_2^{(L)}\alpha_s^3}{\pi^2n^2a}\;
\left[\log\dfrac{na\mu}{2}+\psi(n+l+1)\right];\equn{(3.2c)}$$
$$\eqalign{\delta^{(1)}_{V^{(LL)}}E_{nl}=-\dfrac{C_F\beta_0^2\alpha_s^3}{4\pi^2n^2a}\,
\Big\{\log^2\dfrac{na\mu}{2}+2\psi(n+l+1)\log\dfrac{na\mu}{2}\cr
+\psi(n+l+1)^2+\psi'(n+l+1)\cr
+\theta(n-l-2)\dfrac{2\Gammav(n-l)}{\Gammav(n+l+1)}
\sum^{n-l-2}_{j=0}\dfrac{\Gammav(2l+2+j)}{j!(n-l-j-1)^2}\Big\};\cr}
\equn{(3.2d)}$$
$$\delta^{(1)}_{V_{\rm s.rel}}E_{nl}=\dfrac{C_Fa_2\alpha_s^2}{m}\;
\dfrac{1}{n^3(2l+1)a^2}.\equn{(3.2e)}$$
We recall that constants are collected in the Appendix.
 For the vector states ($\Upsilonv,\,\Upsilonv',\,\Upsilonv'';\;J/\psi,\,\psi',\dots$) 
one has to add the hyperfine shift, at tree level,
$$\delta^{(1)}_{V_{\rm spin}}E_{nl}=\delta_{s1}\delta_{l0}
\dfrac{8C_F\alpha_s}{3n^3m^2a^3}.\equn{(3.2f)}$$
The calculation of the second order 
contribution of $V_1^{(L)}$,  $\delta^{(2)}_{V_1^{(L)}}E_{nl}$, is nontrivial,
 and may be found in the Appendix. We define
$$\delta^{(2)}_{V_1^{(L)}}E_{nl}\equiv 
-m\dfrac{C_F^2\beta_0^2\alpha_s^4}{4n^2\pi^2}
\left\{N_0^{(n,l)}+N_1^{(n,l)}\log \dfrac{na\mu}{2}+
\tfrac{1}{4}\log^2\dfrac{na\mu}{2}\right\} \equn{(3.3a)}$$
and then one has, for the lowest states, 
$$\eqalign{N_1^{(1,0)} = -{\gamma_E \over 2} \simeq -0.288608 \cr
N_1^{(2,0)} = {1 -2 \gamma_E \over 4} \simeq -0.0386078 \cr
N_1^{(2,1)} = {5 -6 \gamma_E \over 12} \simeq 0.128059 \cr
N_0^{(1,0)}= {3 +3 \gamma_E^2 -\pi^2 +6 \zeta(3) \over 12} \simeq 0.111856 \cr
N_0^{(2,0)}=-\tfrac{5}{16}-{\gamma_E \over 4}+{\gamma_E^2 \over 4} - 
{\pi^2 \over 12} + \zeta(3) \simeq 0.00608043 \cr
N_0^{(2,1)}=-\tfrac{865}{432}-{5 \gamma_E \over 12}+{\gamma_E^2 \over 4} - 
{11\pi^2 \over 36} + \zeta(3) \simeq 0.0314472. \cr}\equn{(3.3b)}$$

In addition to this one has to consider the nonperturbative (NP) energy splittings. 
The dominant ones are associated with the gluon condensate and are\ref{6}
$$\eqalign{\delta_{\rm NP}E_{nl}=
m\epsilon_{nl}n^2\pi\langle\alpha_sG^2\rangle\left(\dfrac{na}{2}\right)^4=
m\dfrac{\epsilon_{nl}n^6\pi\langle\alpha_sG^2\rangle}{(mC_F\tilde{\alpha}_s)^4};\cr
\epsilon_{10}=\tfrac{1\,872}{1\,275},\;\epsilon_{20}=\tfrac{2\,102}{1\,326},\;
\epsilon_{21}=\tfrac{9\,929}{9\,945}.\cr}\equn{(3.4)}$$

Because $\langle \alpha_sG^2\rangle\sim\Lambdav^4$, this is of order 
$(\Lambdav/m)^4$ albeit with large coefficients: for all terms we have 
a fourth power of $\alpha_s$ in the denominator, and for $n>1$ the 
$n^6$ in the numerator of \equn{(3.4)} grows very quickly out of hand. In fact it 
is the size of this term that limits the range of validity of our type of 
{\sl ab initio} calculation.

\brochuresubsection{3.2 Higher corrections}
Besides the corrections reported in the previous subsection there are 
a few pieces of the higher order corrections that are known. First of
 all we have the relativistic, $O(v^2)$ corrections to the one loop potential\ref{4}.
 These produce corrections of 
higher order, $\alpha_s^5$, but they are logically independent of the {\sl three loop} ones 
that would produce terms of the same order but, presumably, smaller because of the
 extra $1/\pi^2$ characteristic of radiative corrections. These corrections may be
 incorporated and then can be 
considered to give an indication of the error committed in neglecting 
higher order {\sl perturbative} corrections. They produce, for the ground state, 
the energy shift (ref.~4; typos corrected in ref.~8),
$$\eqalign{\delta^{(1)}_{1\,{\rm loop},\,v^2}E_{10}=A_5+A_S,\cr
A_5=2m\Bigg\{-\dfrac{3C_F^4\beta_0}{32\pi}\left[\log\dfrac{\mu}{mC_F\tilde{\alpha}_s}-
\tfrac{1}{3}-\gammae\right]\cr
-\dfrac{3C_F^4}{16\pi}\left(a_1+
\dfrac{\beta_0\gammae}{2}\right)\cr
-\dfrac{C_F^4a_3}{16\pi}\left[\log\dfrac{1}{C_F\tilde{\alpha}_s}-1\right]+
\dfrac{C_F^4[a_5-(\tfrac{5}{6}+\log\bar{n})a_4]}{16\pi}\Bigg\}\alpha_s^2\tilde{\alpha}_s^3,\cr
A_S=2m\delta_{s1}\dfrac{C_F^4}{6\pi}
\Bigg\{\dfrac{\beta_0}{2}\left(\log\dfrac{na\mu}{2}-\sum_1^n\dfrac{1}{k}-\dfrac{n-1}{2n}\right)\cr
-\dfrac{21}{4}\left(\log\dfrac{n}{C_F\tilde{\alpha}_s}-\sum_1^n\dfrac{1}{k}-\dfrac{n-1}{2n}\right)
+B\Bigg\}\alpha_s^2\tilde{\alpha}_s^3.}
\equn{(3.5)}$$

Next we have higher order NP corrections. These include finite size corrections, 
estimated in ref.~7, and contributions of higher dimensional operators, some of 
which were evaluated in ref.~9. The last produce the shifts,
$$\eqalign{\delta_{\rm NP,\,higher}= { -1 \over m^5 (C_F \tilde{\alpha}_s)^8} h(n,l) O_6 \cr
h(1,0)= {{141912051712}\over {844421875}}\,,
 \quad h(2,0)= {{484859657191424}\over {2040039729}}, \cr
h(2,1)= {{102150951135870976}\over {765014898375}}, \cr
O_6 (\mu)=
{1 \over 108}\left\{ {2^6 \over 3} \pi^2 \alpha_s(\mu) \kappa + 
{3 \over 4} \langle G^3 \rangle
\right\},\quad
\kappa = \alpha_s \langle 0 \vert{\bar q} q \vert 0 \rangle^2,\cr} \equn{(3.6)}$$
and may be used to estimate the size of the higher order NP contributions. For the 
quark condensate the vacuum saturation approximation is assumed, and the value 
of $\kappa$ is taken from 
ref.~10. For $ \langle G^3 \rangle$ one takes the value $0.065\;\gev^6$. Anyway, these 
quantities are poorly known. 

It is important to realize that both (3.5), (3.6) 
should be taken as {\sl indications}. With respect to the first, there is no guarantee 
that the coefficient of the three loop correction is not so large that it offsets the factors of 
$1/\pi$; indeed, this already happens to two loops where the coefficient
 is large, $b_1\simeq 24$. With respect to (3.6), and apart from the fact that it does not 
include all the higher dimensional operators (those associated with size 
corrections are neglected\fnote{The reason for doing so is 
that, at least nominally, the contribution of operators associated with the size, 
$\langle(\partial G(0))^2\rangle$, 
is of higher  order in $\alpha_s$ than the ones considered in (3.6). 
See ref.~9 for details.}), it is clear that one cannot 
consider rigorously a contribution $O(\Lambdav^6/m^6)$ so long as the radiative 
corrections to the $O(\Lambdav^4/m^4)$ terms are not known. Nevertheless, we 
consider (3.5) and  (3.6) as very useful for estimating the theoretical uncertainties 
of our calculation.

\brochuresection{4 Numerical Results}
Using the formulas deduced above one can evaluate the spectrum of heavy 
quarkonium systems. In principle one should take $m,\,\Lambdav,\,\langle\alpha_sG^2\rangle$ from 
other sources and predict the masses of the quarkonium states. In practice 
it is better to use the known masses of the 
lowest states ($\Upsilonv$ and $J/\psi$) to {\sl evaluate} 
the quark masses. The reason is that this produces by far the more
 precise evaluation available at present 
 of these parameters, especially in the case of the $b$ quark. The 
other parameters we take from independent sources. For the QCD parameter $\Lambdav$ 
we take, throughout this section,
$$\Lambdav(n_f=4,\,\hbox{three loops})=0.23^{+0.08}_{-0.05}\;\gev\;
\left[\;\alpha_s(M_Z^2)\simeq0.114^{+0.06}_{-0.04}\;\right],
\equn{(4.1a)}$$ 
and for the gluon condensate, very poorly known,
$$\langle\alpha_sG^2\rangle=0.06\pm0.02\;\gev^4.\equn{(4.1b)}$$

Another matter to be discussed is the choice of the renormalization 
point, $\mu$. As our equations (3.2, 3) show, a {\sl natural} value for this 
parameter is 
$$\mu=\dfrac{2}{na},\equn{(4.1c)}$$
for states with the principal quantum number $n$, and this will be our choice. For 
states with $n=1$ the results of the 
calculation will turn out to depend very little on the value of 
$\mu$, provided it is reasonably close to (4.1c). Higher states are 
another matter; we will discuss our choices when we consider them.
\brochuresubsection{4.1 The 10 state of $\bar{b}b$ and the mass $m_b$}
As stated, we select, for the $\Upsilonv$ state, $\mu=2/a$. We then use \equs~(3.1-4) 
to obtain the values of the $b$ quark mass. To make apparent the contribution 
of the higher corrections, we have performed calculations taking into account only 
$O(\alpha_s^2)$, $O(\alpha_s^3)$, $O(\alpha_s^3)+|v|\times\hbox{1 loop}$ terms, 
and finally the full $O(\alpha_s^4)$ evaluation. The results are reported in Table I below\fnote{We
 have arranged in Table I the results in terms of powers of $\alpha_s$; we could 
have arranged them in increasing number of loops. Cf. \sect~5 for this.}, 
where the errors correspond to the errors in \equs~(4.1a, b).
\setbox1=\vbox{\hsize=15.5truecm\petit 
\smallskip
$$\matrix{&O(\alpha_s^2)&O(\alpha_s^3)&1\,{\rm loop}+{\rm rel.}^*&O(\alpha_s^4)\cr
\mu^2\;(\gev^2)&3.233&4.940&\simeq 2.5&7.019\vphantom{\dint^a}\cr
m_b\;(\gev)&4.752&4.858&4.939^*&
5.015^{+0.101}_{-0.064}\,(\Lambdav)\;\mp0.005\,(\langle\alpha_sG^2\rangle)\;
^{-0.027}_{+0.041} \;(\hbox{vary}\; \mu^2\;{\rm by}\,25\%)
\vphantom{\dint^a}\cr
\bar{m}_b(\bar{m}_b^2)=&4.209&4.307&4.382^*&
4.453^{+0.029}_{-0.015}\,(\Lambdav)\,\mp0.004\,(\langle\alpha_sG^2\rangle)\;
^{-0.025}_{+0.038} \;(\hbox{vary}\; \mu^2\;{\rm by}\,25\%)\vphantom{\dint^a}\cr}$$
\medskip
\centerrule{3cm}
\medskip
\centerline{{\sc Table I}: determinations of the $b$ quark mass
 with increasing accuracy.}
\centerline{$^*$This value is taken from ref.~4, extrapolated to $\Lambdav=0.23\,\gev,\,
\langle\alpha_sG^2\rangle=0.06\,\gev^4$.}
\medskip}
\medskip
\centerline{\boxit{\box1}}
\medskip
In the estimate of the errors, the condition $\mu=2/a$ is 
maintained satisfied when varying $\Lambdav$ while for the error 
due to the variation of $\mu$ the other parameters are kept fixed (i.e., one no more 
has then $\mu=2/a$). The dependence of $m_b$ on $\mu$ should 
be taken as an indication of the theoretical 
uncertainty of our calculation.   
To estimate other theoretical errors in our evaluation we proceed 
as follows. We either calculate $m_b$ including the 
full $O(v^2)$ corrections to one loop, \equs~(3.5). Then we get,
$$(1-{\rm loop}+{\rm rel.})+(2-{\rm loop,\;NR}):
\,t=7.009,\;m_b=5.010;\;\bar{m}_b(\bar{m}_b^2)=4.448;\equn{(4.2a)}$$
or we may include the contribution of higher dimensional NP effects, 
as in (3.6). Then,
$$\hbox{with higher NP effects:}\;m_b=5.018;\;\bar{m}_b(\bar{m}_b^2)=4.455.\equn{(4.2b)}$$

We consider that the best result is that of $O(\alpha_s^4)$ reported 
in Table~I, and take the difference with the quantities given in \equs~(4.2) as 
a further estimation of the theoretical error of the calculation. In this 
way we get our best estimate,
$$\eqalign{m_b=5.015^{+0.101}_{-0.064}\,(\Lambdav)\;\mp0.005\,(\langle\alpha_sG^2\rangle)\;
^{-0.027}_{+0.041} \;(\hbox{vary}\; \mu^2\;{\rm by}\,25\%)
\;\pm 0.006\;({\rm other\; th.\;uncertainty})\cr
\bar{m}_b(\bar{m}_b^2)=4.453^{+0.029}_{-0.016}\,(\Lambdav)
\,\mp0.005\,(\langle\alpha_sG^2\rangle)\;
^{-0.027}_{+0.040} \;(\hbox{vary}\; \mu^2\;{\rm by}\,25\%)
\;\pm0.005\;({\rm other\; th.\;uncertainty}).\cr}
\equn{(4.3)}$$
The values of $\alpha_s(\mu^2)$, $\tilde{\alpha}_s(\mu^2)$ corresponding to 
$\mu^2=7.019\,\gev^2$ are 
$$ \alpha_s(\mu^2)=0.24\;, \tilde{\alpha}_s(\mu^2)=0.40.$$
The ``theoretical" error coming from 
higher dimensional operators and higher order perturbative terms, \equs~(4.2), 
 is comfortably 
smaller that the errors due to the uncertainty on $\Lambdav,\;\langle\alpha_sG^2\rangle$.
 We will henceforth omit these errors, so as not to double count them,
 and consider that the theoretical 
error is only that due to varying $\mu^2$ by 25\%. If we now
 compose all the errors quadratically, then we obtain the estimate reported in 
the Introduction, \equs~(1.1).  

\brochuresubsection{4.2 $M(\Upsilonv)-M(\eta_b)$; the decay
 $\Upsilonv\rightarrow e^+e^-$}
The evaluation of refs.~4, 8 for the hyperfine splitting, and the decay of
 the $\Upsilonv$ into $e^+e^-$ does not change, except that the favoured values of 
$\Lambdav,\,\langle\alpha_sG^2\rangle$ and  
$m_b$ are now somewhat  different. This improves slightly  the agreement 
with experiment for the decay rate. The expressions are,
$$\eqalign{M(V)-M(\eta)=m\,\dfrac{C_F^4\alpha_s(\mu^2)\tilde{\alpha}_s(\mu^2)^3}{3}
\left[1+\delta_{\rm wf}+\delta_{\rm NP}\right]^2\cr
\times\left\{1+\left[\dfrac{\beta_0}{2}\left(\log\dfrac{a\mu}{2}-1\right)
+\tfrac{21}{4}\left(\log C_F\tilde{\alpha}_s+1\right)+B\right]\dfrac{\alpha_s}{\pi}
+\tfrac{1\,161}{8\,704}\,
\dfrac{\pi\langle\alpha_s G^2\rangle}{m^4\tilde{\alpha}_s^6}\right\};\cr}\equn{(4.4a)}$$
$$\eqalign{\Gammav(\Upsilonv\rightarrow e^+e^-)=\Gammav^{(0)}\times\,
\left[1+\delta_{\rm wf}+\delta_{\rm NP}\right]^2\,(1+\delta_{\rm rad}),\cr
\Gammav^{(0)}=2\left[\dfrac{Q_b\alpha_{\rm QED}}{M(\Upsilonv)}\right]^2
\left(mC_F\tilde{\alpha}_s(\mu^2)\right)^3;\cr
\delta_{\rm rad}=-\dfrac{4C_F\alpha_s}{\pi};\;\delta_{\rm wf}=
\dfrac{3\beta_0}{4}\left(\log\dfrac{a\mu}{2}-\gammae\right)\dfrac{\alpha_s}{\pi};
\cr}\equn{(4.4b)}$$
$$\eqalign{
\delta_{\rm NP}=\tfrac{1}{2}
\left[\tfrac{270\,459}{108\,800}+\tfrac{1\,838\,781}{2\,890\,000}\right]
\dfrac{\pi\langle\alpha_s G^2\rangle}{m^4\tilde{\alpha}_s^6}.\cr}$$
The corrections here are fairly large, particularly the radiative
 correction\ref{12} $\delta_{\rm rad}$. Because of this the calculation is
 less reliable than what one would have expected. 
With the values of $m_b$ found in (4.3), one has the numerical results,
$$M(\Upsilonv)-M(\eta)=48.4^{+13.1}_{-4.8}\,(\Lambdav)\,^{+4.9}_{-4.6}\,
(\langle\alpha_sG^2\rangle)\,
^{+7.2}_{-10.2}\,(\mu^2=7.019\pm25\%)\equn{(4.5)}$$
and
$$\Gammav(\Upsilonv\rightarrow e^+e^-)=1.135^{+0.15}_{-0.02}\,(\Lambdav)\,
\pm0.11\,(\langle\alpha_s G^2\rangle)\,^{+0.19}_{-0.27}\,(\mu^2=7.019\pm25\%).\equn{(4.6)}$$
Note that, when varying $\Lambdav,\;\langle \alpha_sG^2\rangle$, 
we have varied $m_b$ according to \equn{(4.3)}, but we have {\sl not} 
varied $m_b$ when varying $\mu$.

 Higher order NP corrections due to the higher dimensional operators 
 introduced in \equn{(3.6)} are also known for the decay rate (see ref. 9). They read
$$
\delta_{NNP}=
{w(n) \over 2 m^6 (C_F {\tilde \alpha})^{10}} \,O_6, \quad \quad
w(1)= -\tfrac{1670626488940208128}{485563688671875}. $$
Size corrections, however, are not known now. $\delta_{NNP}$ 
would produce a shift in the decay rate of $\sim0.11\;\kev$,  
 smaller than the contribution of $\langle\alpha_s G^2\rangle$
 or the  uncertainty caused by e.g. 
varying $\mu$ as in \equn{(4.3)}. We do {\sl not} include 
 $\delta_{NNP}$ either in the evaluation or the error estimate.

The result for the decay is in reasonable agreement with experiment,
$$\Gammav_{\rm exp.}(\Upsilonv\rightarrow e^+e^-)=1.320\pm0.04\,\kev.$$
Composing the errors we obtain the figures quoted in the 
Introduction, \equs~(1.2, 3).

\brochuresubsection{4.3  Higher states ($n=2$) of $\bar{b}b$}
The masses of the states with $n=2$ will be next determined. As is 
clear from the expressions (3.2, 3) the natural choice of scale is now 
$\mu=1/a$ which gives $\mu^2=3.05\,\gev^2$.  If we take this, adding or subtracting 
a 25\% to estimate the dependence of the calculation on the choice of scale then we obtain  
 the results
$$
\eqalign{M(20,\;\hbox{th})-M(20,\;\hbox{exp})=293^{+286}_{-299}\;\mev\;
(\mu^2=3.048\pm25\%),\cr
M(21,\;\hbox{th})-M(21,\;\hbox{exp})=174^{+191}_{-203}\;\mev\;(\mu^2=3.048
\pm25\%).\cr}\equn{(4.7)}
$$
 
We only present the errors that follow from variation of the scale $\mu^2$ 
by 25\%; slightly smaller ones are produced by the errors of 
$\Lambdav,\;\langle\alpha_s G^2\rangle$. We do not explicit this: because of the size 
of the errors in \equn{(4.7)} there is no point in going for 
a more detailed error analysis. 
 
Although they have decreased from th one loop 
evaluations (e. g. ref.~4),the errors are still fairly large here;  
 within them, there is compatibility 
between theory and experiment. Agreement to a few \mev\ for both states 
is obtained if choosing $\mu^2=0.75/a\simeq 2.3\,\gev^2$ or keeping 
$\mu=1/a$ and taking $\langle\alpha_sG^2\rangle=0.036\,\gev^4$: unlike 
for the states with $n=1$ we have now strong dependence of the results 
on the parameters of 
the calculation. This is due to the {\sl large} size of the 
corrections, perturbative and (especially) nonperturbative. This last 
is made more apparent when considering contributions of higher dimensional 
operators\ref{9}, which get completely out of hand for $nl=20$ and 
are very large for $nl=21$. In this 
context, it is satisfactory to realize that it is for this last state (21) 
for which agreement with experiment is best, and theoretical errors smaller.

We will not discuss here the spin and tensor splittings among the states with 
$nl=21$. The inclusion of three loop corrections 
to the potential only affects their calculation in that the preferred value 
for $m_b$ will be different now, which is a minute effect compared with 
the uncertainties of the calculation: one should 
realize\ref{8} that, while the NP corrections to the energy levels with 
principal quantum number $n$ contain a coefficient $n^6/\tilde{\alpha}_s^4$, 
wave functions at the origin get a factor  $\sim n^8/\tilde{\alpha}_s^6$. This 
of course is what makes the calculation of $M(\Upsilonv)-M(\eta_b)$ and the decay
 $\Upsilonv\rightarrow e^+e^-$ much less reliable than that of $M(\Upsilonv)$ 
(or, equivalently, $m_b$) and what makes the evaluation 
of tensor and spin splittings with $n=2$ somewhat marginal. All one can do 
here is {\sl fit} $\mu$ to the data; this is 
the procedure followed in ref.~8, and we have nothing 
new to report on this.  
 
\brochuresubsection{4.4 The 10 state of $\bar{c}c$ and the mass $m_c$}
The value of the parameter $\Lambdav$ used now, corresponding to that in \equn{(4.1a)}, is
$$\Lambdav(n_f=3,\,\hbox{three loops})=0.30^{+0.09}_{-0.05}\;\gev.$$
The values for the $c$ quark mass, deduced from the $J/\psi$ mass 
are then,
$$\eqalign{
O(\alpha_s^4):\;&t=2.623,\;m_c=1.884,\;\bar{m}_c(\bar{m}_c^2)=1.547\cr
 (1-{\rm loop}+{\rm rel.})+(2-{\rm loop,\;NR}):\;& 
t=2.611,\;m_c=1.875;\;\bar{m}_c(\bar{m}_c^2)=1.539.\cr
 {\rm \; with \; higher\; NP \;effects}: \;&
 t=2.634,\;m_c=1.891;\;\bar{m}_c(\bar{m}_c^2)=1.554 .\cr}
\equn{(4.8)}$$
Including errors we obtain the best estimate, analogous to that in (4.3) for the
$b$ quark:
$$\eqalign{m_c=1.884^{+0.157}_{-0.092 }\,(\Lambdav)\;\mp0.012\,(\langle\alpha _sG^2\rangle)\;
^{-0.096}_{+0.156} \;(\hbox{varying}\; \mu^2\;{\rm by}\,25\%)\;\pm\;0.011\;
({\rm th.\;uncertainty})\cr
\bar{m}_c(\bar{m}_c^2)=
1.547^{+0.086}_{-0.049 }\,(\Lambdav)\,\mp 0.011\,(\langle\alpha_sG^2\rangle)\;
^{-0.089}_{+0.145} \;(\hbox{varying}\; \mu^2\;{\rm by}\,25\%)\;\pm\;0.010\;
({\rm th.\;uncertainty}).\cr}
\equn{(4.9)}$$
As is obvious from these equations, the errors are now much larger than for the $b$ quark case, 
but our determination of $m_c$ still competes in accuracy with those based on 
QCD sum rules. 

\brochuresection{5 Discussion}
The calculations of this paper are rather straightforward; but there are a few points 
that merit further discussion. First of all, our values for the quark masses are 
somewhat larger than existing estimates based on sum rules; for the \msbar\ masses,
of $100\sim300\;\mev$, cf. ref.~10. In our opinion this is due to the influence 
of the terms of order $\alpha_s^3,\;\alpha_s^4$ which we take into account, but which the 
sum rule evaluations, that stop at $O(\alpha_s^2)$, do not. Thus we consider our 
estimates to be the more precise and reliable ones.

Secondly, and from Table~I, it may appear that the series is {\sl diverging}: from 
the $O(\alpha_s^2)$ to the $O(\alpha_s^3)$ evaluation, $m_b$ increases by $106\,\mev$ 
but from the last to the $O(\alpha_s^4)$ the increase is of $157$. Actually, convergence 
is reasonably good. The increase between $O(\alpha_s^3)$ and  $O(\alpha_s^4)$ is 
due to two {\sl independent} factors: inclusion of the two 
loop corrections to the potential, responsible 
for $46\;\mev$, and the relativistic corrections. Of these, $64\;\mev$  for tree level corrections 
and $40\;\mev$ for the mixed one loop-velocity correction. Each of the effects is 
small. Thus, if we included velocity corrections at every loop, 
the variation from zero to one to two loops 
would be {\sl diminishing}. This is apparent if we compare the 
value obtained in ref.~4 (corrected for the 
increased values of $\Lambdav,\;\langle\alpha_sG^2\rangle$ we are using now) with our results, 
with a variation of only $60\;\mev$. One can see this more clearly if
 we arrange the calculation in increasing 
number of loops including, at every step, the pertinent relativistic
 corrections\fnote{Nevertheless, 
it is true that the two loop correction is large: $b_1\sim 24,\,b_1/\pi^2\sim2.5$ and
 $\alpha_s^2b_1/\pi^2\sim0.14$: to be compared with $a_1/\pi\sim0.5$, 
$\alpha_s a_1/\pi\sim0.11$. We are 
clearly near the limit of convergence of the perturbative series.} or including 
all loop corrections, but 
in increasing order of the velocity corrections:
$$m_b=\cases{\hbox{tree level incl. rel. correct's.}:\;4.758\cr
\hbox{one loop; incl. rel. correct's.}:\; 4.893\cr
\hbox{full calculation}\;:\;5.015;\cr}$$

$$m_b=\cases{\hbox{static, 2 loop}:\;4.962\cr
\hbox{full calc.}:\;5.015\cr}$$
Finally, we remark the satisfactory stability 
we now have against variation of the renormalization scale, $\mu$. 
This stability of the results against changes of $\mu$ is made apparent by the fact 
that even multiplying or dividing the central value $\mu^2=7.019$ by a factor 
of two only alters the central value of $m_b=5.015\,\gev$ by $98\,\mev$. The stability is due 
mostly to the inclusion of two loop effects; but attention 
should also be paid to the stabilizing influence of the NP corrections. These corrections  
are larger for 
larger $\mu$, exactly the opposite to what happens to the perturbative corrections. One 
could even fix optimal values of $\mu$ as those where the combined perturbative-NP effects 
would show a minimal dependence on $\mu$. This is essentially the procedure 
adopted in ref.~4. Here, and because 
of our much smaller dependence on $\mu$, 
we need not have recourse to such methods.   

\vfill\eject
\brochuresection{Appendix}
\brochuresubsection{Constants}
$$\eqalign{\beta_0=11-\tfrac{2}{3}n_f;\beta_1=102-\tfrac{38}{3}n_f\cr
\beta_2=\tfrac{2847}{2}-\tfrac{5033}{18}n_f+\tfrac{325}{54}n_f^2\cr}$$
$$\eqalign{a_1=&\dfrac{31C_A-20T_Fn_f}{36}\simeq 1.47;\;a_2=\dfrac{C_F-2C_A}{2}\simeq-2.33;\cr
b_1=&\tfrac{1}{16}
\Big\{\left[\tfrac{4343}{162}+6\pi^2-\tfrac{1}{4}\pi^4+\tfrac{22}{3}\zeta(3)\right]C_A^2\cr
&-\left[\tfrac{1798}{81}+\tfrac{56}{3}\zeta(3)\right]C_AT_Fn_f-
\left[\tfrac{55}{3}-16\zeta(3)\right]C_FT_Fn_f+\tfrac{400}{81}T_F^2n_f^2\Big\}\cr
&\simeq 24.30; \cr}$$
$$c_2^{(L)}=a_1\beta_0+\tfrac{1}{8}\beta_1+\tfrac{1}{2}\gammae\beta_0^2.$$
$$\eqalign{a_3=\dfrac{14C_F-21C_A}{3},\;a_4=\dfrac{-16C_F+4C_A}{3},\cr
a_5=2C_F+\dfrac{8C_A}{3}-\dfrac{64T_F}{15}+4T_F\log 2;\cr}$$
$$B=\tfrac{3}{2}(1-\log 2)T_F-\tfrac{5}{9}T_Fn_f+\dfrac{11C_A-9C_F}{18}\simeq 0.29. $$

\brochuresubsection{Second order contribution}
We now calculate $\delta^{(2)}_{V_1^{(L)}}E_{nl}$. 
A simple, and rather accurate approximation may be obtained with
 use of the following trick. The 
second order shift given by a potential $W$ may be written as
$$\delta^{(2)}_{W}E_{nl}=\langle nl|W|\delta^{(1)}R_{nl}\rangle,$$
where
$$|\delta^{(1)}R_{nl}\rangle=P_{nl}|R^{(1)}_{nl}\rangle;$$
$P_{nl}$ is the projector orthogonal to the state $nl$ and
$$|R^{(1)}_{nl}\rangle=|nl\rangle+
 \sum_{n'\neq n}\dfrac{1}{E_n-E_{n'}}\,|n'l\rangle\langle n'l|W|nl\rangle$$
is the wave function to first order. The trick is to use for this 
not the result of a Rayleigh-Schr\"odinger formula, but that 
obtained from a variational principle. For 
our case (cf. ref.~4, \equs~(79-81)), 
$$R^{(1)}_{nl}=\dfrac{2}{n^2b^{3/2}}\sqrt{\dfrac{(n-l-1)!}{(n+l)!}}
\left(\dfrac{2r}{nb}\right)\ee^{-r/nb}L^{2l+1}_{n-l-1}(2r/nb),$$
$$b^{-1}=a^{-1}\left[1+\dfrac{\log(na\mu/2)+\psi(n+l+1)-1}{2\pi}\beta_0\alpha_s\right].$$
 For the ground state, this gives, after a trivial calculation,
$$\delta^{(2)}_{V^{(L)}_1}E_{10}=
-m\dfrac{\beta_0^2C^2_F\alpha_s^3\tilde{\alpha}_s}{16\pi^2}\left[\log^2\dfrac{a\mu}{2}
-\gammae\right]^2.$$
This simple method gives correctly the coefficients of $\log a\mu,\;\log^2a\mu$ 
and misses the constant term by $\sim 10\%$.

 For the {\sl exact} calculation
 one uses the representation of the Coulombic Green's function given e.g. 
 by Voloshin\fnote{An estimate neglecting the
 continuum contribution was been given in ref. 14. The approximation is not good, although 
since the quantity is small its effect in the evaluation of 
$m_b$ was not important.}. We write
$$\delta_{V_1^{(L)}}^{(2)} E_{nl}= 
\sum_{k \not= n} {|\langle n,l| V_1^{(L)} |k,l\rangle|^2 \over E_n -E_k},
$$
and the sum over $k$ includes an 
integral over the continuous part of the spectrum.
Instead of doing this computation directly we have used the more general function 
$$\eqalign{\sum_{k} {|\langle n,l| V_1^{(L)} |k,l\rangle|^2 \over E -E_k}= 
{|\langle n,l| V_1^{(L)} |n,l\rangle |^2 \over E -E_n}+
\sum_{k \not= n} {|\langle n,l| V_1^{(L)} |k,l\rangle|^2 \over E -E_k}\cr
= {A_{(-1)} \over E -E_n} + A_{(0)} +
A_{(1)}\,(E -E_n)+\ldots}
$$
and then $ \delta_{V_1^{(L)}}^{(2)} E_{nl}= A_{(0)}$. For the function defined above 
one uses the representation of the Coulombic Green's function given e.g.
 by Voloshin in the second article in ref. 6 [Note that 
there is a misprint in formula (15) there, and $(s+l+1)!$ must be
 changed to $(s+2\,l+1)!$]. In this way we get for the different $N_{nl}$ defined 
in \equn{(3.3)}, and for arbitrary quantum numbers,
$$\eqalign{N_1^{(n,l)} = {\psi(1 + l + n)-1 \over 2};\qquad 
N_0^{(n,l)} = {1 \over 4}\psi(1 + l + n)\big[\psi(1 + l + n) -2\big] \cr
+{n \over 2} \left[
{ (n-l-1)! \over (n+l)!}
\sum^{n-l-2}_{s=0} {(s + 2\,l + 1)! \over s!\,(s + l + 1 - n)^3 } +
{ (n+l)! \over (n-l-1)!}
\sum^{\infty}_{s=n-l} {s! \over (s+2\,l+1)!\,(s+l+1-n)^3 } 
\right].}
$$
\vfill\eject
\brochuresection{Acknowledgments}
The authors are grateful to J. Soto, who collaborated 
in the early stages of this paper, for interesting discussions. One of us (A.P.) 
is also indebted to the Generalitat de Catalunya for financial 
support (CIRIT, contract GRQ93-1047).

\brochuresection{References}
\item{1}{A. Billoire, Phys. Lett. {\bf B92} (1980) 343.}
\item{2}{W. Buchm\"uller, Y. J. Ng and S.-H. H. Tye, Phys. Rev. {\bf D24} 
(1981) 3003.}
\item{3}{S. N. Gupta and S. Radford, Phys. Rev. {\bf D24} (1981) 2309 and (E) 
{\bf D25} (1982) 3430; S. N. Gupta, S. F. Radford 
and W. W. Repko, {\sl ibid.} {\bf D26} (1982) 3305.}
\item{4}{S. Titard and F. J. Yndur\'ain, Phys. Rev. {\bf D49} (1994) 6007.}
\item{5}{M. Peter, Phys. Rev. Lett. {\bf 78} (1997) 602.}
\item{6}{M. B. Voloshin, Nucl. Phys. {\bf B154} (1979) 365 and Sov. J. Nucl. Phys. 
{\bf 36} (1982) 143; H. Leutwyler, Phys. Lett. {\bf B98} (1981) 447.}
\item{7}{Yu. A. Simonov, S. Titard and  F. J. Yndur\'ain, Phys. Lett. {\bf B354} (1995) 435.}
\item{8}{S. Titard and F. J. Yndur\'ain, Phys. Rev. {\bf D51} (1995) 6348;
the analysis corrected for some states  
by A. Pineda,  Phys. Rev. {\bf D55} (1997) 407.}
\item{9}{A. Pineda, Nucl. Phys, {\bf B494} (1997) 213.}
\item{10}{S. Narison, Phys. Lett. {\bf B341} (1994) 73 and 
Acta Phys. Pol., {\bf B26} (1995) 687;  M. Jamin and A. Pich, hep-ph 9702276.}
\item{11}{R. Coquereaux, Phys. Rev. {\bf D23} (1981) 1365 to one loop;
N. Gray et al., Z. Phys. {\bf C48} (1990) 673 to two loops.}
\item{12} {E. Braaten, S. Narison and A. Pich, Nucl. Phys. {\bf B373} (1992) 581.}
\item{13}{R. Barbieri et al., Phys. Lett. {\bf 58B} (1975) 455; {\sl ibid.} Nucl. 
Phys. {\bf B154} (1979) 535.}
\item{14}{F. J. Yndur\'ain, Proc. QCD97 (hep-ph/9708448).}

\bye